\documentstyle[preprint,aps]{revtex}

\begin{document}

\draft


\tighten

\title{Symmetry Origin of Infrared Divergence Cancellation in 
Topologically Massive Yang-Mills Theory}

\author{ W.F. Chen\renewcommand{\thefootnote}{\dagger}\footnote{E-mail: wchen@theory.uwinnipeg.ca}\renewcommand{\thefootnote}{\star}\footnote{Address after September 1: Department of Mathematics and Statistics, University of Guelph, Guelph, Ontario, Canada N1G 2W1} 
}

\address{Department of Physics, 
 University of Winnipeg, Winnipeg, Manitoba, R3B 2E9 Canada \\
 and \\
Winnipeg Institute for Theoretical Physics, Winnipeg, Manitoba}

\maketitle

\begin{abstract}
\noindent 
We manifestly verify that the miraculous cancellation of the 
infrared divergence of the topologically massive Yang-Mills theory 
in the Landau gauge is completely determined by a new vector symmetry 
existing in its large topological mass limit, the Chern-Simons theory. 
Furthermore, we show that the cancellation theorem proposed by 
Pisarski and Rao is an inevitable consequence of this new vector symmetry. 
 
\vspace{3ex}

\end{abstract}

\vspace{3ex}

There had been an upsurge in the perturbative investigation
of 2+1-dimensional non-Abelian Chern-Simons field theory
following a seminal work by Witten\cite{ref1}, in which
a miraculous relation between quantum Chern-Simons theory
and two-dimensional conformal invariant WZW model was found. 
Various regularization schemes were utilized in the perturbative
calculation\cite{ref2,ref3,ref4,ref5,ref6,ref7,ref8}. Now looking back  
what remarkable features have been revealed by the perturbative
theory, it seems to us that there are three points. In addition to 
the famous finite renormalization of the gauge coupling, one is 
an explicit confirmation on an earlier assumption by Jackiw\cite{ref9} 
that the Chern-Simons theory is the large topological mass limit of the 
topologically massive Yang-Mills theory (TMYM) \cite{ref7,ref10,ref11},
in which the Chern-Simons term can provide a topological mass to the 
Yang-Mills field without recourse to the Higgs mechanism\cite{ref12}; 
The other is the exposure of a new vector symmetry existing only in 
the Landau gauge fixed Chern-Simons theory\cite{ref13}, which similar 
to the BRST symmetry puts the gauge and ghost fields in a nonlinear 
multiplet and is thus entitled the Landau vector supersymmetry. It was 
verified that this new symmetry has non-trivial dynamical effects:
it protects the Chern-Simons theory from getting quantum 
correction\cite{ref14,ref14a}. In particular, it is intimately related 
to the ambiguity of the finite renormalization of the gauge coupling 
constant: if the regularization method adopted in a concrete perturbative 
calculation preserves the BRST symmetry but violates the Landau vector 
supersymmetry, the coupling constant receives
a finite renormalization\cite{ref15}, while reversely the coupling
constant will keep its classical value\cite{ref16}. There exists no 
regularization schemes that can preserve both of the Landau vector 
supersymmetry and the BRST symmetry simultaneously\cite{ref15,ref17}.  
 
It was well known that in the Landau gauge TMYM is ultraviolet 
finite \cite{ref12,ref18,ref19} and its beta function vanishes 
identically. Consequently, there is no dynamical generated mass 
scale and the form factor of a Green function can only be the 
form of $f(p^2/m^2)$ with $p$ being certain external 
momentum\footnote{The mass dimension of 
the coupling constant can be removed by a re-definition of the fields, 
so the only massive parameter left is the topological mass\cite{ref7}.}. 
Thus the fact that the Chern-Simons theory is the large topological mass 
limit of TMYM means that it can be defined as the 
infrared limit of TMYM. An equivalent statement is that TMYM is infrared 
finite. On the other side, the Landau vector supersymmetry arises in 
the Landau gauge-fixed Chern-Simons theory. In view of these two aspects, 
we cannot help thinking that the Landau vector supersymmetry, which 
arises in the low-energy limit of TMYM, probably has some connection 
with the cancellation of the infrared divergence. The aim of this letter 
is to give an explicit verification. 

The cancellation of the infrared divergence in TMYM 
was investigated in detail in a beautiful paper 
by Pisarski and Rao\cite{ref19}. They first observed that in the 
one-loop vacuum polarization tensor, the infrared divergences come 
from the gauge parameter dependent part of the gluon propagator, 
the ghost loop amplitude, and the combination of the parity-odd pieces 
from each three-gluon vertex 
and the gluon propagator in the gluon loop contribution.
However, under the choice of the Landau gauge, the gauge dependent part
vanishes, the combination of parity-odd pieces in the gluon loop amplitude
and the ghost loop contribution unexpectedly cancel in the infrared limit. 
The same phenomenon happens for the one-loop three-gluon vertex. 
Based on these observations, Pisarski and Rao concluded that 
the Chern-Simons part is completely responsible for the 
cancellation of the infrared divergences in TMYM, despite that their 
ultraviolet behaviour is greatly different. More concretely, 
the cancellation of the infrared divergence
is determined by the odd parity property of the gauge-fixed 
Chern-Simons theory, and only in the Landau gauge the Chern-Simons 
theory can present this feature. This conclusion render them to 
formulate a cancellation theorem to explain the miraculous
cancellation of the infrared divergence in TMYM\cite{ref19}: 
In Landau gauge the Chern-Simons theory has only a finite 
renormalization and its quantum effective action 
takes the same form as the classical one. The cancellation of
infrared divergences contributed by the ghost field and the parity-odd 
parts of the gauge field  is entirely dominated by this theorem.

Despite that this theorem gets  the essence of the infrared divergence 
cancellation, but it looks far from being regarded as
the origin of a dynamical phenomenon. Especially, 
the cancellation
takes place at the level of the Green function, 
this theorem is not clear enough to show how the cancellation occurs among the Green functions. There
must be certain symmetry hidden behind this empirical theorem, which will
relate various Green functions through the Ward identities, 
to implement the cancellation
of the infrared divergence. Thus we place the hope on the Landau vector
supersymmetry to play such a role.

The classical action of TMYM in Euclidean space-time is\cite{ref12}
\begin{eqnarray}
S&=&-im\int d^3x\epsilon_{\mu\nu\rho}
\left(\frac{1}{2}A_{\mu}^a{\partial}_{\nu}A_{\rho}^a
+\frac{1}{3!}gf^{abc}A_{\mu}^aA_{\nu}^bA_{\rho}^c\right) 
+\frac{1}{4} \int d^3x F_{\mu\nu}^aF^a_{\mu\nu},
\label{eq1}
\end{eqnarray}
where it is assumed that $m>0$. To remove the mass dimension of
the gauge coupling so that it is explicit that the Chern-Simons theory 
can be regarded as the large topological mass limit of TMYM at 
classical level, 
we rescale the field and the coupling,
$A{\rightarrow}m^{-1/2}A$ and $g{\rightarrow}m^{-1/2}g$. Consequently,
the classical action (\ref{eq1}) becomes\cite{ref7}
\begin{eqnarray}
S&=&-i\int d^3x\epsilon_{\mu\nu\rho}
\left(\frac{1}{2}A_{\mu}^a{\partial}_{\nu}A_{\rho}^a
+\frac{1}{3!}gf^{abc}A_{\mu}^aA_{\nu}^bA_{\rho}^c\right)
+\frac{1}{4m} \int d^3x F_{\mu\nu}^aF^a_{\mu\nu}.
\label{eq2}
\end{eqnarray}
In the Lorentz gauge-fixing condition, the gauge-fixing and the ghost 
field parts of the total classical effective action are as the following,
\begin{eqnarray}
S_{g}=\int d^3x \left[-B^a\partial_\mu A_\mu^a-\partial_{\mu}\bar{c}^a
\left(\partial_{\mu}c^a+gf^{abc}A_{\mu}^bc^c\right)
+\frac{\xi}{2}(B^a)^2\right].
\label{eq3}
\end{eqnarray}
As in the usual gauge theory, the gauge-fixed action has the well known
BRST symmetry\cite{ref13},
\begin{eqnarray}
\delta A_{\mu}^a=D_{\mu}^{ab}c^b,~~\delta c^a=-\frac{g}{2}f^{abc}c^bc^c,~~
\delta\bar{c}^a=B^a,~~\delta B^a=0.
\label{eq4}
\end{eqnarray}
However, in the Landau gauge choice, $\xi=0$, the large topological mass limit
of TMYM,  the gauge-fixed
Chern-Simons theory has a new BRST-like vector symmetry
\cite{ref13,ref14,ref14a,ref15},
\begin{eqnarray}
V_{\mu} A_\nu^a=i\epsilon_{\mu\nu\rho}
\partial_{\rho}c^a,~~V_\mu B^a=-(D_\mu c)^a, ~~
V_{\mu}  c^a=0, ~~V_{\mu}  \bar{c}^a=A_\mu^a.
\label{eq5}
\end{eqnarray}

At quantum level,  
writing out formally the generating functional
with the inclusion of the external sources for the fields
and their variations under the supersymmetry (\ref{eq5}), one can derive
the Ward identities corresponding to the Landau vector supersymmetry
of the Chern-Simons theory in a standard way\cite{ref14,ref14a}.
In the topologically massive Yang-Mills theory this symmetry is explicitly
broken by the Yang-Mills term, but the broken Ward identity can still be
written out\cite{ref15,ref16}. Here we take a shortcut to 
derive the needed identities\cite{ref15}. Consider a general functional 
$F[\Phi]$ of the field $\Phi=\left(A_\mu^a,B^a,c^a,\bar{c}^a\right)$, 
the invariance of the quantum observable
\begin{eqnarray}
\langle F[\Phi]\rangle =\int {\cal D}{\Phi}F[\Phi]e^{-S[\Phi]}
\end{eqnarray}
under an arbitrary infinitesimal transformation $\Phi\rightarrow\Phi
+\delta\Phi$ yields the following identity,
\begin{eqnarray}
\left\langle \frac{\partial F[\Phi]}{\partial \Phi}\delta\Phi 
-\delta S[\Phi]F[\Phi]\right\rangle =0.
\end{eqnarray}
 Choosing the functions $F[\Phi]=A_\mu^a (x)\bar{c}^b(y)$ and 
 $F[\Phi]=A_\mu^a (x)A_\nu^b(y)\bar{c}^c(z)$ in
TMYM, respectively, the Landau vector supersymmetry
transformation (\ref{eq5}) leads to
\begin{eqnarray}
\left\langle A_\mu^a (x)A_\nu^b (y)\right\rangle
=i\epsilon_{\mu\nu\rho}\partial^x_\rho\left\langle 
c^a(x)\bar{c}^b(y)\right\rangle +\left\langle A_\mu^a (x)\bar{c}^b(y)
V_\nu S_{YM}\right\rangle,
\label{eq8}
\end{eqnarray}
and 
\begin{eqnarray}
\left\langle A_\mu^a (x)A_\nu^b (y)A_\rho^c(z)\right\rangle
&=&i\epsilon_{\mu\rho\lambda}\partial^x_\lambda\left\langle 
c^a(x)A_\nu^b(y)\bar{c}^c(z)\right\rangle
+ i\epsilon_{\nu\rho\lambda}\partial^y_\lambda\left\langle 
A_\mu^a(x)c^b(y)\bar{c}^c(z)\right\rangle \nonumber\\
&&+\left\langle A_\mu^a (x)A_\nu^b(y)\bar{c}^c(z)
V_\rho S_{YM}\right\rangle.
\label{eq9}
\end{eqnarray}
A straightforward calculation gives\cite{ref15} 
\begin{eqnarray}
V_\mu S_{YM}&=& {\cal O}_\mu^{(0)}+{\cal O}_\mu^{(1)}+{\cal O}_\mu^{(2)};\\
{\cal O}_\mu^{(0)}&=&-\frac{i}{m}\int d^3x\epsilon_{\mu\nu\rho}
\partial_{\rho}c^a\Box A^a_{\nu};\\
{\cal O}_\mu^{(1)}&=&\frac{i}{m}gf^{abc}\int d^3x\epsilon_{\mu\nu\rho}
\partial_\rho c^a
\left[-2 A_\lambda^b\partial_{\lambda}A^c_{\nu}+
A_\lambda^b\partial_\nu A^c_{\lambda}
-\left(\partial_\lambda A_\lambda^b\right) A^c_{\nu}\right];\\
{\cal O}_\mu^{(2)}&=&\frac{i}{m}g^2f^{abc}f^{ade}
\int d^3x\epsilon_{\mu\nu\rho}A_\lambda^b \partial_{\rho}c^c 
A^d_{\lambda} A^e_{\nu}.
\end{eqnarray}
Eq.(\ref{eq8}) shows that the broken Landau vector supersymmetry
has established a direct relation 
between the gauge field and the ghost field propagators.
In momentum space, it reads
\begin{eqnarray}  
D_{\mu\nu}(p)=\epsilon_{\mu\nu\rho} p_{\rho} S(p)
+D_{\mu\rho}(p)\Omega_{\rho\nu}(p)S(p),
\label{eq14}
\end{eqnarray}
where $D_{\mu\nu}(p)$ and $S(p)$ 
are the propagators of gauge and ghost fields in momentum space, respectively; 
the composite vertex $\Omega (p)$ at tree level comes from
 ${\cal O}^{(0)}_{\mu}$. 
Eq.(\ref{eq14}) formally leads to
\begin{eqnarray}
\epsilon_{\mu\nu\rho}p_\rho D^{-1}_{\lambda\mu}(p) S(p)
=\delta_{\lambda\nu}-\Omega_{\lambda\nu}(p) S(p).
\label{eq18}
\end{eqnarray}
Eq.(\ref{eq9}) relates the three-point function
of the gauge field to the three-point function of the gauge field
and the ghost fields. In momentum space it is 
\begin{eqnarray}
D_{\mu\alpha}(p)D_{\nu\beta}(q)D_{\rho\gamma}(r)
\Gamma_{\alpha\beta\gamma}^{abc}(p,q,r)
&=&-\epsilon_{\mu\rho\lambda}p_{\lambda}S(p)D_{\nu\beta}(q)S(r)
\Gamma_{\beta}(p,r,q)\nonumber\\
&&-\epsilon_{\nu\rho\lambda}q_{\lambda}D_{\mu\alpha}(p) S(q)S(r)
\Gamma_{\alpha}(q,r,p)\nonumber\\
&&+D_{\mu\alpha}(p)D_{\nu\beta}(q)S(r)
\widetilde{\Gamma}_{\alpha\beta\rho}^{abc}(p,q,r),\nonumber\\
 p+q+r&=&0.
\label{eq18a}
\end{eqnarray}
where $\Gamma_{\mu\nu\sigma}^{abc}$ and $\Gamma_{\mu}^{abc}$
are the three-gluon vertex and the ghost-ghost-gluon vertex,
respectively; $\widetilde{\Gamma}^{abc}_{\mu\nu\rho}(p,q,r)$ 
at tree-level comes from the composite operator ${\cal O}_\nu^{(1)}$.
Extracting the one particle irreducible (1PI) part, we obtain
a relation between the three-gluon vertex and the ghost-ghost-gluon 
vertex, 
\begin{eqnarray}
D_{\rho\sigma}(r)
\Gamma_{\mu\nu\sigma}^{abc}(p,q,r)&=&-\epsilon_{\alpha\rho\lambda}
p_{\lambda}D_{\mu\alpha}^{-1}(p)S(p)S(r)
\Gamma_{\nu}^{acb}(p,r,q)\nonumber\\
&&-\epsilon_{\alpha\rho\lambda}
q_{\lambda}D_{\nu\alpha}^{-1}(q)S(q)S(r)
\Gamma_{\nu}^{bca}(q,r,p)+ S(r)\widetilde{\Gamma}^{abc}_{\mu\nu\rho}(p,q,r),
\nonumber\\ 
 p+q+r&=&0.
\label{eq19}
\end{eqnarray}
The identities (\ref{eq14}) and (\ref{eq19}) at tree-level can be easily
verified with the bare propagators and vertices.
The substitution of (\ref{eq18}) into (\ref{eq19}) further yields
\begin{eqnarray}
D_{\rho\sigma}(r)
\Gamma_{\mu\nu\sigma}^{abc}(p,q,r)&=&
\delta_{\mu\rho}S(r)\Gamma_\nu^{abc}(p,r,q)
-\delta_{\nu\rho}S(r)\Gamma_\mu^{abc}(q,r,p)\nonumber\\
&&-\Omega_{\mu\rho}(p)S(p)S(r)\Gamma_\nu^{abc}(p,r,q)
+\Omega_{\nu\rho}(q)S(q)S(r)\Gamma_\mu^{abc}(q,r,p)\nonumber\\
&&+S(r)\widetilde{\Gamma}^{abc}_{\mu\nu\rho}(p,q,r).
\label{eq23}
\end{eqnarray}

In the following, we shall show explicitly how the identity (\ref{eq23}),
born out of the Landau vector supersymmetry, enforces the cancellation
of infrared divergence. Consider the vacuum polarization tensor contributed 
from the skeleton diagrams of the ghost and the gluon loops,
we have
\begin{eqnarray}
\Pi_{\mu\nu}^{ab}(p)&=&
\frac{1}{2}\int \frac{d^3k}{(2\pi)^3}\Gamma_{\mu\lambda\rho}^{acd}(p,k,-k-p)
D_{\lambda\beta}(k)\Gamma_{\nu\alpha\beta}^{bdc}(-p,k+p,-k)D_{\alpha\rho}(k+p)
\nonumber\\
&&-\int \frac{d^3k}{(2\pi)^3}\Gamma_\mu^{acd}(p,k,-k-p)S(k)
\Gamma_\nu^{bdc}(-p,k+p,-k)S(k+p).
\label{eq24}
\end{eqnarray}
Despite that Eq.(\ref{eq24}) is written as the form of one-loop Feynman 
diagram amplitude, it can actually be regarded as
a vacuum polarization tensor at any 
higher order since the insertion of the vertices and propagators 
in these two one-loop skeletons can be chosen to be any order 
as desired. The first term in the R.H.S. of Eq.(\ref{eq24}) 
is the contribution from the skeleton of gluon loop
and the second one from the ghost loop skeleton, these are the only 
two sources of the infrared divergence\cite{ref19}. 
Substituting Eq.(\ref{eq23}) into Eq.(\ref{eq24}), we obtain
\begin{eqnarray}
\Pi_{\mu\nu}^{ab}(p)&=&
\frac{1}{2}\int \frac{d^3k}{(2\pi)^3}\left\{S(-k-p)S(-k)
\left[\delta_{\mu\alpha}\Gamma_\lambda^{acd}(p,-k-p,k)
-\delta_{\lambda\alpha}\Gamma_\mu^{acd}(k,-k-p,p)\right.\right.\nonumber\\
&&-\Omega_{\mu\alpha}(p)S(p)\Gamma_\lambda^{acd}(p,-k-p,k)
+\Omega_{\lambda\alpha}(k)S(k)\Gamma_\mu^{acd}(k,-k-p,p)\nonumber\\
&&+\left.\widetilde{\Gamma}^{acd}_{\mu\lambda\alpha}(p,k,-k-p)\right]
\left[\delta_{\nu\lambda}\Gamma_\alpha^{bdc}(-p,-k,k+p)
-\delta_{\lambda\alpha}\Gamma_\mu^{bdc}(k+p,-k,-p)\right.\nonumber\\
&&-\Omega_{\nu\lambda}(-p)S(-p)\Gamma_\alpha^{bdc}(-p,-k,k+p)
+\Omega_{\lambda\alpha}(k+p)S(k+p)\Gamma_\nu^{bdc}(k+p,-k,-p)\nonumber\\
&&+\left.\left.\widetilde{\Gamma}^{bdc}_{\mu\alpha\lambda}(-p,k+p,-k)\right]
\right\}\nonumber\\
&-&\int \frac{d^3k}{(2\pi)^3}\Gamma_\mu^{acd}(p,k,-k-p)S(k)
\Gamma_\nu^{bdc}(-p,k+p,-k)S(k+p) \nonumber\\
&=&\frac{1}{2}\int \frac{d^3k}{(2\pi)^3}S(-k)S(-k-p)
\left[\Gamma_\mu^{bdc}(-p,-k,k+p)\Gamma_\nu^{acd}(p,-k-p,k)\right.\nonumber\\
&&-\Gamma_\mu^{acd}(p,-k-p,k)\Gamma_\nu^{bdc}(k+p,-k,-p)
-\Gamma_\mu^{acd}(k,-k-p,p)\Gamma_\nu^{bdc}(-p,-k,k+p)\nonumber\\
&&\left.+3\Gamma_\mu^{acd}(k,-k-p,p)\Gamma_\nu^{bdc}(k+p,-k,-p)\right]
\nonumber\\
&&+\frac{1}{2}\int \frac{d^3k}{(2\pi)^3}
 S(-k)S(-k-p)\left\{\left[\delta_{\mu\alpha}\Gamma_\lambda^{acd}(p,-k-p,k)
-\delta_{\lambda\alpha}\Gamma_\mu^{acd}(k,-k-p,p)\right]\right.\nonumber\\
&&\times \left[\Omega_{\lambda\alpha}(k+p)S(k+p)\Gamma_\nu^{bdc}(k+p,-k,-p)
-\Omega_{\nu\lambda}(-p)S(-p)\Gamma_\alpha^{bdc}(-p,-k,k+p)\right]\nonumber\\
&&+\left[\Omega_{\lambda\alpha}(k)S(k)\Gamma_\mu^{acd}(k,-k-p,p)
-\Omega_{\mu\alpha}(p)S(p)\Gamma_\lambda^{acd}(p,-k-p,k)
\right]\nonumber\\
&&\left.\times
\left[\delta_{\nu\lambda}\Gamma_\alpha^{bdc}(-p,-k,k+p)
-\delta_{\alpha\lambda}\Gamma_\nu^{bdc}(k+p,-k,-p)\right]\right\}
\nonumber\\
&&+\frac{1}{2}\int \frac{d^3k}{(2\pi)^3}
 S(-k)S(-k-p)\left\{\left[\delta_{\mu\alpha}\Gamma_\lambda^{acd}(p,-k-p,k)
-\delta_{\lambda\alpha}\Gamma_\mu^{acd}(k,-k-p,p)\right]
\right.\nonumber\\
&&\times \widetilde{\Gamma}_{\nu\alpha\lambda}^{bdc}(-p,k+p,-k)
+\widetilde{\Gamma}_{\mu\lambda\alpha}^{acd}(p,k,-k-p)
\left[ \delta_{\nu\lambda}\Gamma_\alpha^{bdc}(-p,-k,k+p)\right.
\nonumber\\
&&\left.\left.-\delta_{\lambda\alpha}
\Gamma_\nu^{bdc}(k+p,-k,-p)\right]\right\}
\nonumber\\
&+& \frac{1}{2}\int \frac{d^3k}{(2\pi)^3}S(-k)S(-k-p)\left\{\left[
\Omega_{\lambda\alpha}(k)S(k)\Gamma_\mu^{acd}(k,-k-p,p)\right.
\right.\nonumber\\
&&\left.-\Omega_{\mu\alpha}(p)S(p)\Gamma_\lambda^{acd}(p,-k-p,k)\right]
\widetilde{\Gamma}^{bdc}_{\nu\alpha\lambda}(-p,k+p,-k)\nonumber\\
&&+\widetilde{\Gamma}^{acd}_{\mu\lambda\alpha}(p,k,-k-p)
\left[\Omega_{\lambda\alpha}(k+p)S(-k)\Gamma_\nu^{bdc}(k+p,-k,-p)\right.
\nonumber\\
&&\left.\left.-\Omega_{\nu\lambda}(-p)S(-p)\Gamma_\alpha^{bdc}(-p,-k,k+p)
\right]\right\}\hspace{1cm} 
\nonumber\\
&+&\frac{1}{2}\int \frac{d^3k}{(2\pi)^3}
S(-k)S(-k-p)\left\{\left[\Omega_{\lambda\alpha}(k)S(k)
\Gamma_\mu^{acd}(k,-k-p,p)
\right.\right.\nonumber\\
&&\left.-\Omega_{\mu\alpha}(p)S(p)\Gamma_\lambda^{acd}(p,-k-p,k)\right]
\left[\Omega_{\lambda\alpha}(k+p)S(k+p)\Gamma_\nu^{bdc}(k+p,-k,-p)\right.
\nonumber\\
&&\left.\left.-\Omega_{\nu\lambda}(-p)S(-p)\Gamma_\alpha^{bdc}(-p,-k,k+p)
\right]\right\} 
\nonumber\\
&+&\frac{1}{2}\int \frac{d^3k}{(2\pi)^3}
 S(-k)S(-k-p)\widetilde{\Gamma}^{acd}_{\mu\lambda\alpha}(p,k,-k-p)
\widetilde{\Gamma}^{bdc}_{\mu\alpha\lambda}(-p,k+p,-k)
\nonumber\\
&-&\int \frac{d^3k}{(2\pi)^3}S(k)S(k+p)\Gamma_\mu^{acd}(p,k,-k-p)
\Gamma_\nu^{bdc}(-p,k+p,-k). 
\label{eq25}
\end{eqnarray}
The first term  in the expansion
of Eq.(\ref{eq25}) is only associated with the contribution
from the ghost propagators and the ghost-ghost-gluon vertices, 
and hence it is the term in the gluon loop skeleton that contains 
the infrared divergence\cite{ref19}. The global
gauge symmetry requires that the ghost-ghost-gluon vertex 
at any order must be of the form,
\begin{eqnarray}
\Gamma_\mu ^{abc}(p,q,r)=f^{abc}\Gamma_\mu (p,q,r).
\label{eq26}
\end{eqnarray}
Moreover, the Ward identities from the
BRST symmetry given in Eq.(\ref{eq4}) determines the ghost
field propagator and the ghost-ghost-gluon vertex 
are\cite{ref19,ref20},
\begin{eqnarray}
S(p)&=& -\frac{1}{p^2\left[ 1+\Sigma (p^2/m^2)\right]};\nonumber\\
\Gamma_{\nu}^{abc}(p,q,r)&=& p_{\mu}\gamma_{\mu\nu}^{abc}(p,q,r),
~~~~~~ p+q+r=0,
\label{eq28}
\end{eqnarray}
where $\gamma_{\mu\nu}^{abc}$ is the composite ghost-gluon vertex
function of the following general form, 
\begin{eqnarray}
\gamma_{\mu\nu}^{abc}(p,q,r)&=& f^{abc}\left(\epsilon_{\mu\nu\rho}p_{\rho}A_1
+\epsilon_{\mu\nu\rho}q_{\rho}A_2+\epsilon_{\mu\lambda\rho}p_{\lambda}
q_{\rho} p_\nu A_3
+\epsilon_{\mu\lambda\rho}p_{\lambda}q_{\rho}q_\nu A_4\right.\nonumber\\
&&+\epsilon_{\nu\lambda\rho}p_{\lambda}q_{\rho}p_\mu A_5
+\epsilon_{\nu\lambda\rho}p_{\lambda}q_{\rho}q_\mu A_6
+g_{\mu\nu}A_7+p_\mu p_\nu A_8+p_\mu q_\nu  A_9\nonumber\\
&&\left.+ p_\nu q_\mu A_{10} +q_\mu q_\nu A_{11}\right),\nonumber\\
A_i&{\equiv}&A_i(p^2,q^2,r^2,m), ~~i=1,2, {\cdots}, 11; ~~~p+q+r=0.
\label{eq29}
\end{eqnarray}
Using Eqs.(\ref{eq26})--(\ref{eq29}), we can easily find
that up to a constant coefficient term 
the first term of Eq.(\ref{eq25}) cancels with
the contribution from the ghost loop skeleton (i.e. the last term) 
in the low-energy limit,
\begin{eqnarray}
&&\lim_{p^2\longrightarrow 0}\int \frac{d^3k}{(2\pi)^3}
\frac{1}{2}C_V\delta^{ab}S(-k)S(-k-p)
\left[\Gamma_\mu (p,-k-p,k)\Gamma_\nu (k+p,-k,-p)\right.\nonumber\\
&&+\Gamma_\mu (k,-k-p,p)\Gamma_\nu (-p,-k,k+p)
+2\Gamma_\mu (p,k,-k-p)\Gamma_\nu (-p,k+p,-k)\nonumber\\
&&\left. -\Gamma_\mu (-p,-k,k+p)\Gamma_\nu (p,-k-p,k)
-3\Gamma_\mu (k,-k-p,p)\Gamma_\nu (k+p,-k,-p)\right]=0, 
\label{eq27}
\end{eqnarray}
where $C_V$ is the quadratic Casimir operator in the adjoint
representation of the gauge group, $f^{acd}f^{bcd}=C_V\delta^{ab}$.

To complete the proof that the infrared divergence indeed cancels
in Eq.(\ref{eq25}), we need to show that the remained terms are 
also free from infrared divergence. The remained terms is composed of the
ghost field propagator, the ghost-ghost-gluon vertex and the composite
vertices relevant to the Landau vector supersymmetric variation
of the Yang-Mills term. The identity (\ref{eq14}) yields\footnote{
In fact, Eq.(\ref{eq14}) yields two possibilities, 
$p_{\mu} \Omega_{\mu\nu}(p)=0$ or proportional to $p_{\nu}$,
but the tree level and one-loop results\cite{ref15}, 
$\Omega^{(0)}_{\mu\nu}(p)=-p^2/m\epsilon_{\mu\nu\rho}p_\rho$ and
$\Omega^{(1)}_{\mu\nu}(p)= 3/(4\pi)g^2C_V (p^2\delta_{\mu\nu}-p_\mu p_\nu)$,
exclude the second one.}
\begin{eqnarray}
p_\mu \Omega_{\mu\nu}(p)=0,
\label{eq30}
\end{eqnarray}
and hence $\Omega_{\mu\nu}$ should be the following general form,
\begin{eqnarray}
\Omega_{\mu\nu}(p)=m\epsilon_{\mu\nu\rho} p_{\rho}
B_1\left(\frac{p^2}{m^2}\right)
+(p^2\delta_{\mu\nu}-p_\mu p_\nu)B_2\left(\frac{p^2}{m^2}\right).
\label{eq31}
\end{eqnarray}
As for the composite ghost-gluon-gluon vertex associated with
the Landau vector supersymmetry transformation, 
$\widetilde{\Gamma}_{\mu\nu\rho}(p,q,r)$, the identity (\ref{eq18a})
leads to a relation between $\widetilde{\Gamma}_{\mu\nu\rho}$ and 
the ghost-ghost-gluon vertex,
\begin{eqnarray}
r_{\rho}\widetilde{\Gamma}_{\mu\nu\rho}(p,q,r)=
\epsilon_{\alpha\lambda\rho} p_\lambda q_\rho \left[
S(p)D^{-1}_{\mu\alpha}(p)\Gamma_\nu (p,r,q)
-S(q) D^{-1}_{\nu\alpha}(q)\Gamma_\mu (q,r,p)\right].
\label{eq32}
\end{eqnarray}
Thus the infrared divergence-free of the ghost-ghost-gluon 
vertex $\Gamma_\mu$ implies that $\widetilde{\Gamma}_{\mu\nu\rho}$ 
should also has no infrared divergence. The other argument in favour 
of the infrared divergence-free of $\widetilde{\Gamma}_{\mu\nu\rho}$
is the observation that at tree-level it is relevant to the three-gluon 
vertex of the three-dimensional Yang-Mills theory,
\begin{eqnarray}
\widetilde{\Gamma}_{\mu\nu\rho}^{(0)}(p,q,r)&=&-\frac{i}{m}
\epsilon_{\nu\lambda\sigma}r_{\sigma}\left[g_{\mu\lambda}(p-q)_{\rho}
+g_{\lambda\rho} (q-r)_{\mu}+g_{\rho\mu}(r-p)_{\lambda}\right]
\nonumber\\
&=&-\frac{i}{m}\epsilon_{\nu\lambda\sigma}r_{\sigma}
\Gamma_{(YM)\mu\lambda\rho}^{(0)}(p,q,r),
\label{eq33}
\end{eqnarray}
where $\Gamma_{(YM)\mu\lambda\rho}^{(0)}$ is the tree-level three-gluon 
vertex of the Yang-Mills part of TMYM. There is no infrared divergence 
in three dimensional Yang-Mills theory
due to its superrenormalizability\cite{ref21}. The relation
(\ref{eq33}) will probably be modified at quantum level by quantum 
correction, but we still think that it gives a favourable support on 
the infrared divergence-free of $\widetilde{\Gamma}_{\mu\nu\rho}$. 

The information collected in Eqs.(\ref{eq28}), (\ref{eq29}),
(\ref{eq31}), (\ref{eq32}) and (\ref{eq33}) on propagators
and vertices enable us to use the inductive method
to prove the infrared divergence-free of the remained terms
of Eq.(\ref{eq25}). The procedure is as following.
We first show by concrete calculation that in dimensional
regularization the terms other than the first and the last ones 
in Eq.(\ref{eq25}) have the $p^2{\rightarrow}0$
limit, then assume that this is satisfied at the $n$th order;
After a lengthy analysis we find that at $n+1$ order the remained terms of
Eq.(\ref{eq25}) indeed have no infrared divergence. 
Based on this observation, we conclude that the cancellation of 
the infrared divergence
has occurred in the whole vacuum polarization tensor. 

Similar analysis are applied to the quantum 
three-gluon vertex. It is 
shown that with Eqs.(\ref{eq23}) the infrared divergences coming from
the ghost loop skeleton and the pure $\epsilon$-parts of the gluon loop 
skeleton cancel\footnote{In concrete
perturbative calculation, depending on the choice of a regularization
scheme, Eq.(\ref{eq34}) at one-loop may have a finite term,
$C_1\epsilon_{\mu\nu\rho}$$+C_2/m[\delta_{\mu\nu}(p-q)_\rho
$$+\delta_{\nu\rho}(q-r)_\mu+\delta_{\rho\mu}(r-p)_\mu]$ with $C_1$ 
and $C_2$ being constants.},
\begin{eqnarray}
&&\lim_{p^2,q^2,r^2\rightarrow 0}\left[\Gamma_{(gl)\mu\nu\rho}^{abc}(p,q,r)
+\Gamma_{(gh)\mu\nu\rho}^{abc}(p,q,r)\right]\nonumber\\
&=&\lim_{p^2,q^2,r^2\rightarrow 0}\left\{\int\frac{d^3k}{(2\pi)^3}
\Gamma_{\mu\mu_1\mu_2}^{aa_1a_2}(p,-p-k,k)D_{\mu_2\rho_1}(k)
\Gamma_{\rho\rho_1\rho_2}^{ca_2b_1}(-p-q,-k,p+q+k)\right.\nonumber\\
&&\times D_{\rho_2\nu_1}(p+q+k)
\Gamma_{\nu\nu_1\nu_2}^{bb_1a_1}(q,-p-q-k,p+k)D_{\nu_2\mu_1}(p+k)
\nonumber\\
&-&\int\frac{d^3k}{(2\pi)^3}\left[\Gamma_\mu^{aa_1a_2}(p,-p-k,k)S(k)
\Gamma_{\rho}^{ca_2b_1}(-p-q,-k,p+q+k)S(k+p+q)\right.\nonumber\\
&&\times
\Gamma_\nu^{bb_1a_1}(q,-k-p-q,k+p)S(k+p)
+\Gamma_\mu^{aa_1a_2}(p,k+q,-k-p-q)S(k+q)\nonumber\\
&&\times \left.\left.\Gamma_{\nu}^{bb_1a_1}(q,k,-k-q)S(k)
\Gamma_\rho^{ca_2b_1}(-p-q,k+p+q,-k)S(k+p+q)\right]\right\}
=0,
\label{eq34}
\end{eqnarray}
the three-gluon vertex function is thus infrared finite.
In above equation,  $\Gamma_{(gl)\mu\nu\rho}^{abc}$ and 
$\Gamma_{(gh)\mu\nu\rho}^{abc}$ denote the quantum three-gluon vertex 
contributed by the gluon loop skeleton and the ghost loop skeleton, 
respectively.

Having proved that the Landau vector supersymmetry protects the 
topologically massive Yang-Mills theory from getting the infrared 
divergence, in the following  we shall have a look at the relation
between the Landau vector supersymmetry and the cancellation 
theorem proposed by Pisarski and Rao\cite{ref19}.

First, the infrared cancellation theorem makes it possible for
the Landau vector supersymmetry to persist after the quantum correction.
The main content of Pisarski and Rao's infrared divergence cancellation 
theorem is that in the Landau gauge the Chern-Simons term of TMYM 
keeps its classical form up to a finite wave function
and vertex renormalization, i.e. the quantum effective action of 
Chern-Simons theory is
\begin{eqnarray}
\Gamma_{CS}[A,c,\bar{c},B]&\stackrel{\rm \xi=0}{=}&
-i\int d^3x Z_A\epsilon_{\mu\nu\rho}
\left(\frac{1}{2}A_{\mu}^a\partial_{\nu}A_{\rho}^a+\frac{1}{3!}
\frac{Z_V}{Z_A}g_Rf^{abc}A_{\mu}^aA_{\nu}^bA_{\rho}^c\right)\nonumber\\
&&-\int d^3x \left[B^a\partial_{\mu}A_{\mu}^a+Z_c\partial_{\mu}\bar{c}^a
\left(\partial_\mu c^a+\frac{Z_V^{\prime}}{Z_c}g_Rf^{abc}A_\mu^b c^c\right)
\right]\nonumber\\
&=&-i\int d^3x Z_A\epsilon_{\mu\nu\rho}
\left(\frac{1}{2}A_{\mu}^a\partial_{\nu}A_{\rho}^a+\frac{1}{3!}
Z_A^{1/2}gf^{abc}A_{\mu}^aA_{\nu}^bA_{\rho}^c\right)\nonumber\\
&&-\int d^3x \left[B^a\partial_{\mu}A_{\mu}^a+Z_c\partial_{\mu}\bar{c}^a
\left(\partial_\mu c^a+Z_A^{1/2}gf^{abc}A_\mu^b c^c\right)
\right] ,
\label{eq35}
\end{eqnarray}
where the fields are all renormalized ones; $g$ and $g_R$ are the bare 
and the renormalized gauge couplings, respectively; 
$Z_A$ and $Z_g$  are the wave 
function renormalization constants for the gauge and ghost fields, 
respectively; $Z_V$ and $Z_V^{\prime}$ are the vertex renormalization 
constants for the three-gluon vertex and the ghost-ghost-gluon field vertex, 
respectively. Further, the Slavnov-Taylor identity from the 
quantum BRST symmetry
requires that $Z_V/Z_A=Z_V^{\prime}/Z_c$ \cite{ref19,ref20}.
The form (\ref{eq35}) clearly shows that that the quantum Chern-Simons action 
has the following renormalized Landau vector supersymmetry,
\begin{eqnarray}
V_{R\mu} A_\nu^a &=& iZ_A^{-1/2}Z_c^{1/2}\epsilon_{\mu\nu\rho}
\partial_{\rho}c^a, ~~~ V_{R\mu}c^a=0,~~~ 
V_{R\mu}\bar{c}^a=Z_c^{-1/2}Z_A^{1/2}A_\mu^a, \nonumber\\
V_{R\mu}B^a &=& -Z_A^{1/2}Z_c^{1/2}\partial_\mu c^a
-Z_V^{\prime}Z_A^{1/2}Z_c^{-1/2}g_Rf^{abc}A_{\mu}^b c^c
\nonumber\\
&=&-Z_A^{1/2}Z_c^{1/2}\partial_\mu c^a-Z_VZ_c^{1/2}Z_A^{-1/2}
g_Rf^{abc}A_\mu^bc^c.
\label{eq36}
\end{eqnarray}
Since the Landau vectors supersymmetry transformation changes
parity, so any parity-even terms generated from the quantum correction
such as $(F_{\mu\nu})^2$ etc will break the Landau vector supersymmetry. 
Therefore, the above infrared divergence cancellation theorem guarantees 
that Landau vector supersymmetry is completely anomaly free.

On the other side, there had appeared several different ways 
proving that once the Landau vector 
supersymmetry is imposed, the Ward identity corresponding to it 
will enforce that the quantum Chern-Simons action to be the 
form of (\ref{eq35})\cite{ref14,ref14a}. 
The first method is combining the Landau vector supersymmetry
and the BRST transformation invariance together to form a $N=1$ supersymmetry
algebra, then with respect to this superalgebra, arranging 
the energy-momentum tensor, the ghost number
current, the BRST current and a tensor current corresponding
to the Landau vector supersymmetry into a supermultiplet\cite{ref14}. 
With the fact that the anomalies of the component currents 
in a supermultiplet also constitute a supermultiplet\cite{ref22}, 
it was shown by making use of various
Ward identities that the quantum Chern-Simons theory is scale invariant,
its beta function and the anomalous dimensions of every fields 
vanish identically. This means that the quantum Chern-Simons term keeps 
its classical form\cite{ref22}. The second way is considering 
the Ward identity implied by the Landau vector supersymmetry
and the Schwinger-Dyson equations for the propagators of the 
ghost and gauge fields\cite{ref14a}. This method elegantly 
manifested the vanishing of the radiative correction to the 
Chern-Simons theory\cite{ref14a}. In addition, the concrete 
perturbative calculation in a regularization scheme preserving 
the Landau vector supersymmetry also explicitly confirms that the 
Chern-Simons action indeed receives no quantum correction\cite{ref2}. 
In fact, even one chooses a regularization schemes violating the 
Landau vector supersymmetry in the perturbative Chern-Simons 
theory, one typical example of which is the hybrid regularization
of higher covariant derivative plus dimensional regularization with 
the Yang-Mills action as the higher covariant derivative term, 
the explicit calculation shows that after the regulator is removed 
the theory still keeps its classical form up to a finite 
renormalization of the gauge coupling\cite{ref3,ref4,ref5,ref6,ref7,ref8}. 
Thus the cancellation theorem is an inevitable consequence of
the Landau vector supersymmetry. Based on above arguments considered from 
both sides, we conclude that the existence of Landau vector 
supersymmetry of Chern-Simons theory is equivalent to the 
cancellation cancellation theorem proposed by Pisarski and Rao\cite{ref19}.

 In summary, we have verified that the infrared divergence 
cancellation of topologically massive Yang-Mills theory in the 
Landau gauge originates completely from a new vector symmetry 
in the gauge-fixed Chern-Simons term. 
This is an unusual property of TMYM. 
It is well known that in four-dimensional case,
the infrared divergences arise in a massless field theory. They cancel
out in the scattering cross sections when the degeneracy of a real particle
with massless particle states is considered as summarized by the celebrated 
Kinoshita-Lee-Nauenberg theorem\cite{ref23}, where cancellation
has no explicit relevance with the symmetries of the theory. While here in
three-dimensional TMYM, the cancellation of the infrared divergence can 
adhere to certain symmetry of the theory. Thus it is significant to
point out this exotic feature.

 Furthermore, we have analyzed the relation between the Landau vector
supersymmetry and the infrared divergence cancellation theorem
proposed earlier by Pisarski and Rao. On one hand, 
the cancelation theorem ensures that the Landau vector 
supersymmetry will persist after quantum correction; On the other hand, the
Landau vector supersymmetry determine that infrared divergences are doomed
to cancel. Consequently, Chern-Simons theory can be defined as 
the large topological mass limit of TMYM and Pisarski and Rao's 
cancellation theorem arises. However, there is a difference
between the Landau vector supersymmetry and the cancellation theorem:
the Landau vector supersymmetry has displayed the origin
of the infrared divergence cancellation of TMYM in the Landau gauge, 
while the cancellation theorem has only emphasized the consequence of 
the infrared divergence cancellation .

\acknowledgments

 This work is supported by the Natural Sciences and Engineering Research
Council of Canada. I would like to thank Professor G. Kunstatter
for useful discussions and Professors M. Chaichian  and R. Kobes 
for encouragement and help.

\end{document}